\documentclass[prd,twocolumn,showpacs,amsmath,amssymb]{revtex4}

\usepackage{graphicx}

\begin{document}

\vspace*{0.5cm}

\title{Monojet Searches for MSSM Simplified Models}

\author{Alexandre Arbey$^{1,2}$}
\email{alexandre.arbey@ens-lyon.fr}

\author{Marco Battaglia$^{2,3}$}
\email{marco.battaglia@ucsc.edu}

\author{Farvah Mahmoudi$^{1,2}$,}%
\altaffiliation{Also Institut Universitaire de France, 103 boulevard Saint-Michel, 75005 Paris, France}
\email{mahmoudi@in2p3.fr}

\affiliation{~\\$^1$\mbox{Univ Lyon, Univ Lyon 1, ENS de Lyon, CNRS,}\\
\mbox{Centre de Recherche Astrophysique de Lyon UMR5574, F-69230 Saint-Genis-Laval, France}\vspace*{0.2cm}\\
$^2$CERN, CH-1211 Geneva, Switzerland\vspace*{0.2cm}\\
$^3$University of California at Santa Cruz,\\
Santa Cruz Institute of Particle Physics, CA 95064, USA\vspace*{0.2cm}}

\begin{abstract}

We explore the implications of monojet searches at hadron colliders in the minimal supersymmetric extension of
the Standard Model (MSSM). To quantify the impact of monojet searches, we consider simplified MSSM scenarios with neutralino dark matter. The monojet results of the LHC Run~1 are reinterpreted in the context of several MSSM simplified scenarios, and the complementarity with direct supersymmetry search results is highlighted. We also investigate the reach of monojet searches for the Run~2, as well as for future higher energy hadron colliders.

\end{abstract}

\maketitle


\section{Introduction}

The minimal supersymmetric extension of the Standard Model (MSSM) is the most studied scenario beyond the SM, and the lightest neutralino is a favourite candidate for dark matter (DM), when R-parity is conserved. Searches for events characterised by the emission of a single hard jet, used as a signature of the hard scattering process, are usually considered as a probe of direct production of invisible dark matter particles in $pp$ collisions. The study of monojet signature was pioneered by the Tevatron experiments \cite{Abazov:2003gp,Acosta:2003tz}. Interpretations of the results of direct searches for new particles at the LHC are often performed in the context of simplified scenarios.
Highly constrained SUSY models, such as the Constrained MSSM (CMSSM)~\cite{Kane:1993td,Ellis:2002rp}, have been studied in the past. Currently the attention has shifted towards simplified models \cite{Alwall:2008ag,Alves:2011sq,Alves:2011wf}, where only few degrees of freedom, such as the neutralino mass and the mass splitting to the lightest SUSY particles, define the relevant phenomenology.
In this study, we investigate the implications of monojet searches in the context of the simplified models in the MSSM with neutralino DM by reinterpreting the LHC Run~1 results in a quantitative way, and compare them to the constraints obtained from direct SUSY searches in the jet/lepton + MET channels.  

The lack of signals of low energy SUSY in the LHC Run~1 data already sets strong constraints on the mass spectrum of the SUSY particles, in the constrained MSSM models. More general MSSM scenarios with less ad-hoc universality assumptions, such as the phenomenological MSSM (pMSSM)~\cite{Djouadi:1998di}, where constraints on the colored states do not affect non-colored sparticles, or scenarios with long decay chains or compressed spectra still remain largely viable \cite{Dreiner:2012gx,Dreiner:2012sh,LeCompte:2011cn,LeCompte:2011fh,Bhattacherjee:2012mz,Bhattacherjee:2013wna,Arbey:2013iza}. Compressed scenarios are particularly interesting since the observed dark matter relic density can be achieved in these scenarios thanks to the enhanced effective cross sections due to co-annihilations. In particular, small mass splittings between squarks and gluino with the lightest neutralino can lead to final states with less energetic jets or leptons, thus reducing the detection efficiency and signal acceptance at the LHC. Monojet searches are particularly sensitive to such scenarios, and we shall show that they are indeed powerful in constraining the MSSM, provided all the involved processes are correctly taken into account. Monojet signals in the context of the MSSM have already been discussed in specific
scenarios~\cite{Carena:2008mj,Allanach:2010pp,Drees:2012dd,Dreiner:2012gx,Dreiner:2012sh,Cullen:2012eh,Delgado:2012eu,Arbey:2013aba,Han:2013usa,Arbey:2013iza,Schwaller:2013baa,Low:2014cba}.

The connections of monojet and dark matter searches at the LHC are discussed in section 2. Section 3 describes the way monojet searches can be affected in the MSSM and presents the numerical set-up. The implications of the monojet searches in simplified MSSM scenarios are presented in section 4. Section 5 addresses the sensitivity of the monojet searches at higher center of mass energies and luminosities. The conclusions are given in section 6.

\section{Dark matter searches at the LHC}

\begin{figure}[t!]
\includegraphics[width=3.5cm]{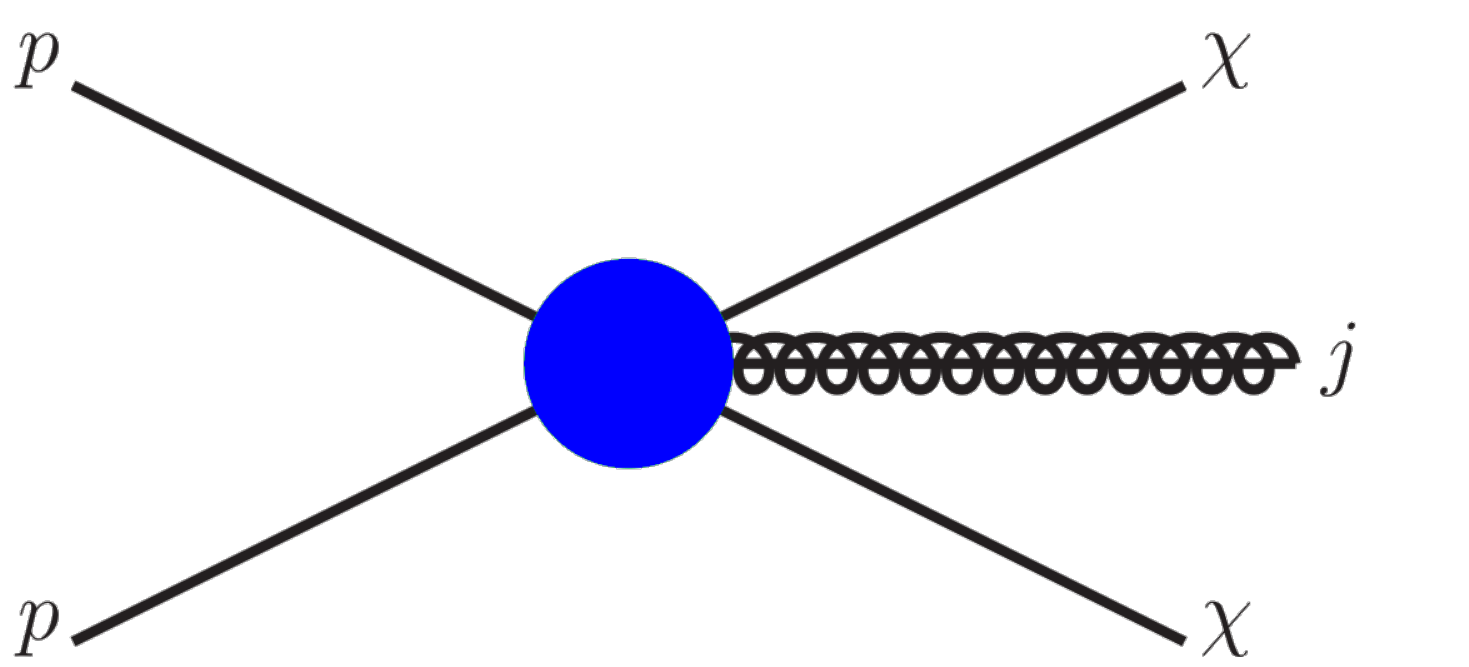}
\caption{Schematic representation of monojet events in the effective or simplified dark matter approaches.\label{fig:standard_monojets}}
\end{figure}

Monojet searches at the LHC consist in looking for events with one high-$p_T$ jet and missing energy, and are therefore particularly well-suited for the search of dark matter particles. A schematic representation of monojet events in simplified scenarios in which a dark matter candidate and a mediator are added to the Standard Model is given in Fig.~\ref{fig:standard_monojets}.

The interpretation of monojet searches has been often performed in the context of effective scenarios, where operators linking quarks and gluons to two DM particles are considered. Hence the constraints obtained on the dark matter particle mass are dependent on the operator under consideration, and can be compared directly to the results of DM direct detection experiments by computing the scattering cross section of DM with standard matter \cite{Goodman:2010yf,Goodman:2010ku}. More recently, the validity of the effective approach has been questioned \cite{Shoemaker:2011vi,Busoni:2013lha,Busoni:2014sya,Busoni:2014haa,Racco:2015dxa}, and instead simplified scenarios with different configurations of dark matter candidates and mediators have been suggested to probe dark matter at the LHC \cite{Abdallah:2015ter,Abercrombie:2015wmb}. The constraints obtained from monojet searches are dependent on the natures, masses and couplings of the dark matter candidates and mediator particles. For a specific set-up, it is possible to reinterpret the results in terms of scattering cross sections of DM with protons and compare it to direct detection experiment results.

\section{Monojet searches in the MSSM}

\begin{figure}[t!]
\includegraphics[width=4.3cm]{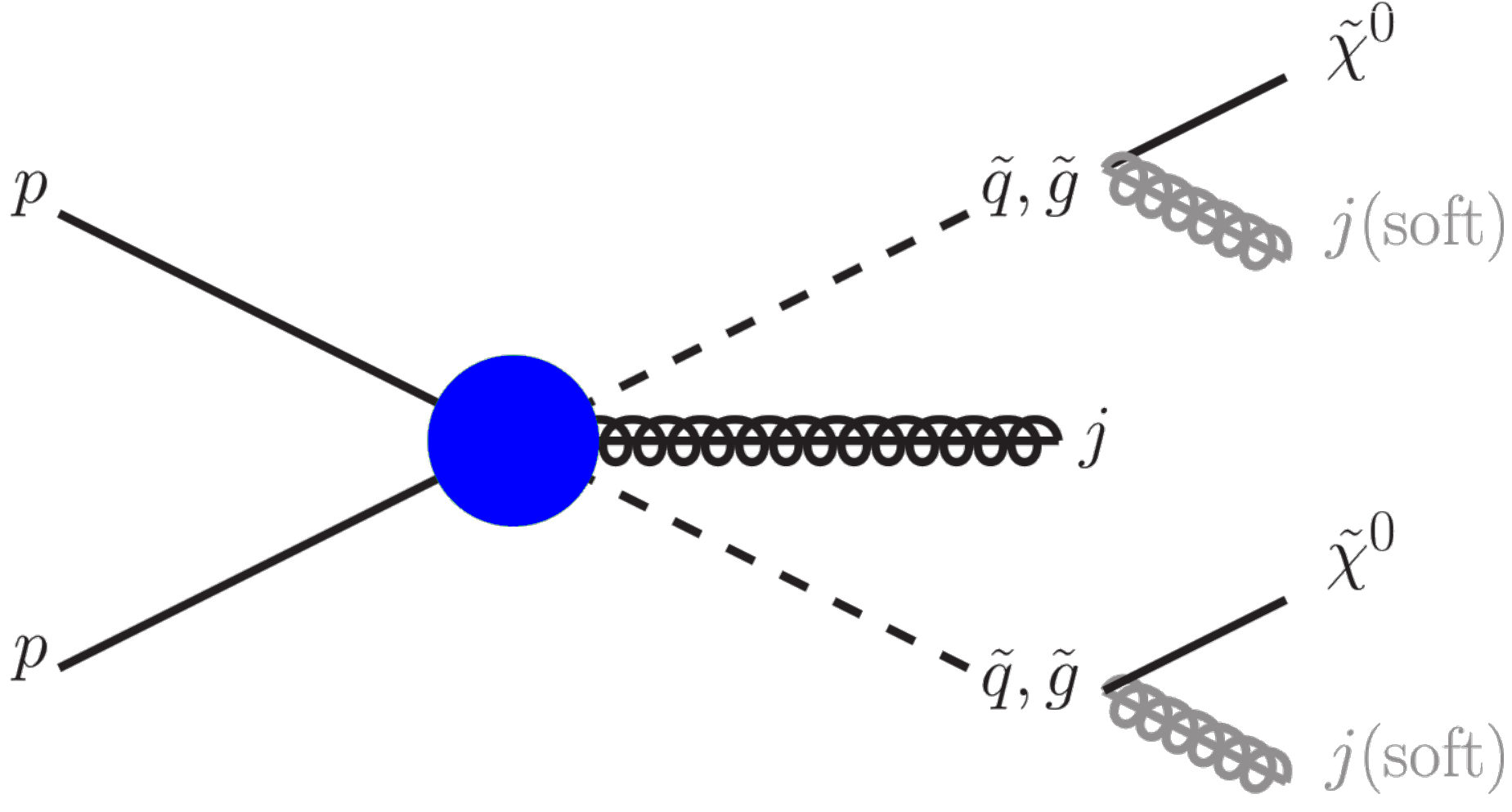}
\caption{Schematic representation of monojet events in the MSSM, where the squarks and gluinos can decay to invisible soft jets and neutralinos.\label{fig:mssm_monojets}}
\end{figure}

In the following we discuss the case of MSSM with R-parity conservation and neutralino dark matter. Monojets in this scenario can be generated by the final states with two neutralinos and one hard jet, as in Fig.~\ref{fig:standard_monojets}, but more importantly two neutralinos, one hard jet and additional soft jets or particles invisible in the detectors. Such final states occur in particular when two squarks or gluinos are produced in addition to a hard jet, as shown in Fig.~\ref{fig:mssm_monojets}. This happens in particular in scenarios with compressed spectra, where the direct SUSY searches are less sensitive but the cross section for the monojet topologies is enhanced by the strong production of degenerate squarks or gluinos. Contrary to the case of simplified or effective approaches, there is no correlation in the MSSM between the monojet production cross section, which can probe the strong sector, and the neutralino scattering cross section with matter, which is sensitive to the electroweak sector, so that monojet searches cannot be considered anymore as dark matter searches, but as complementary channels to the direct SUSY searches and to dark matter cosmological and astrophysical observables.

In this analysis, we use MadGraph 5 \cite{Alwall:2014hca} to compute the full $2\to3$ matrix elements corresponding to all the combinations of $pp \to \tilde{q}/\tilde{g} + \tilde{q}/\tilde{g} + j$, $pp \to \tilde{\ell} + \tilde{\ell} + j$ and $pp \to \tilde{\chi} + \tilde{\chi} + j$, where $\tilde{q}$ refers to a squark of any type and generation, $\tilde{g}$ to the gluino, $\tilde{\ell}$ to any type of slepton, $\tilde{\chi}$ to any electroweakino, and $j$ to a hard jet. Here we do not restrict ourselves to initial state radiation of a monojet. To generate events we adopt the CTEQ6L1 parton distribution functions \cite{Pumplin:2002vw}. Hadronisation is performed using PYTHIA 8 \cite{Sjostrand:2006za,Sjostrand:2007gs}, and detector effects are simulated with DELPHES 3.0 \cite{deFavereau:2013fsa}. 

The exclusion by the monojet searches is assessed based on the ATLAS \cite{Aad:2014nra} and CMS \cite{Khachatryan:2014rra} analyses, using the same cuts, selection efficiencies, acceptances and backgrounds, and predictions for higher energies are obtained by rescaling the background and assessing the sensitivity without modifying the experimental set-up applied in the 8 TeV analyses. In this sense, our analysis is rather conservative as no optimisation is considered. Also, systematic uncertainties have been shown to have an important effect on the limits that can be derived using the monojet signatures \cite{Dreiner:2012sh,Arbey:2013iza,Baer:2014cua}. Here we account for these systematics by adding a 30\% uncertainty on the cross sections.

Signal selection cuts corresponding to each of the analyses are applied to the simulated signal events. The number of SM background events in the signal regions are taken from the estimates reported by the experiments. When the experimental analysis investigates several signal regions, such as the ATLAS and CMS monojet analyses, we calculate the region giving the largest signal-to-background ratio and we only use that region for determining the exclusion. The 95\% confidence level (C.L.) exclusion in presence of background only is determined using the CLs method~\cite{Read:2002hq}.

In addition, we compute the relic density with SuperIso Relic \cite{Arbey:2009gu}, as well as dark matter direct detection observables with MicrOMEGAS \cite{Belanger:2013oya}. We compare the results to the dark matter density measurement of Planck \cite{Ade:2015xua} and to the results of LUX \cite{Akerib:2013tjd} for DM direct detection. Finally, the electroweak observables are computed with a modified version of SuperIso \cite{Mahmoudi:2007vz,Mahmoudi:2008tp}.

\section{MSSM simplified scenarios}
We consider various sets of simplified models and investigate the complementarity with the traditional jets/leptons + MET SUSY searches. These models are characterised by a light neutralino accompanied by at least one additional heavier sparticle, while the other sparticle masses are much heavier. In practice, the light sparticles are lighter than about 1 TeV, and the other masses are adjusted in the range 10--40 TeV in order to obtain a correct light Higgs mass of 125 GeV. The trilinear couplings of the third generation fermions are chosen in order to have no mixing, and $\tan\beta$ is set to 10, an intermediate value which allows us to be consistent with the flavor constraints. The five sets of models correspond to one light neutralino and: a light gluino, degenerate scalar quarks, a light scalar bottom quark, a light scalar top quark, and light neutralino 2 and chargino 1. For the five scenarios, in addition to the discussion of the LHC supersymmetry and monojet searches, we also checked that the $W$ boson mass and electroweak oblique parameters are consistent with the LEP measurements \cite{Schael:2013ita}. Concerning dark matter direct detection, we found that for all these scenarios, the neutralino-nucleon spin-independent scattering cross section is always smaller than $10^{-11}$ pb, which is well below the LUX limits \cite{Akerib:2013tjd}, but within reach of the expected sensitivity of LZ \cite{Akerib:2015cja}. A general feature exhibited by the scenarios we investigated is the significant improvement in sensitivity in the regions with small mass splittings. These regions are especially important since they correspond to the parameter range where coannihilation processes bring the neutralino relic density in agreement with the CMB data, as highlighted by the red lines on our plots.

\subsection{Light gluino scenario}

\begin{figure}[t!]
 \includegraphics[width=8.cm]{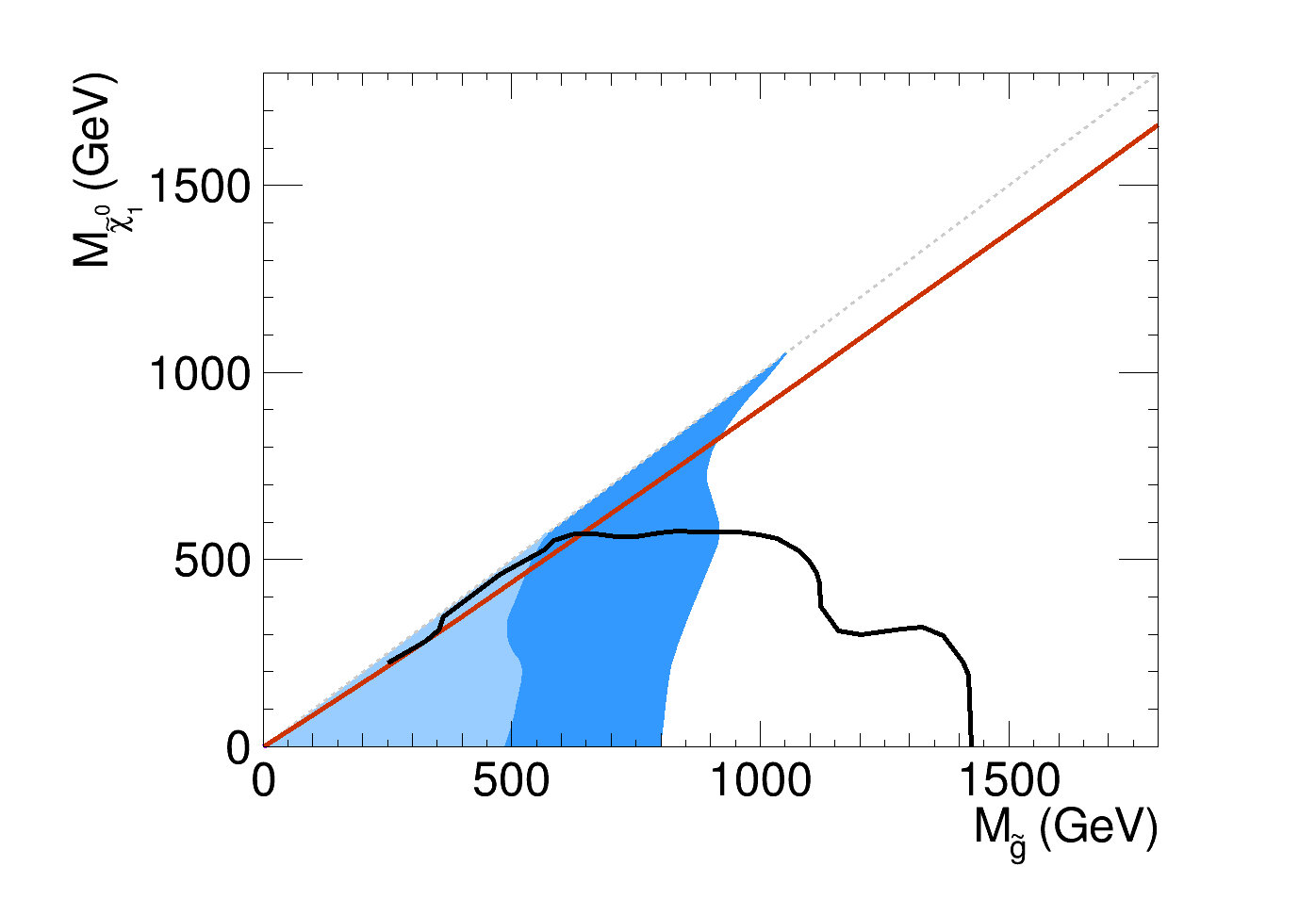}\\
 \includegraphics[width=8.cm]{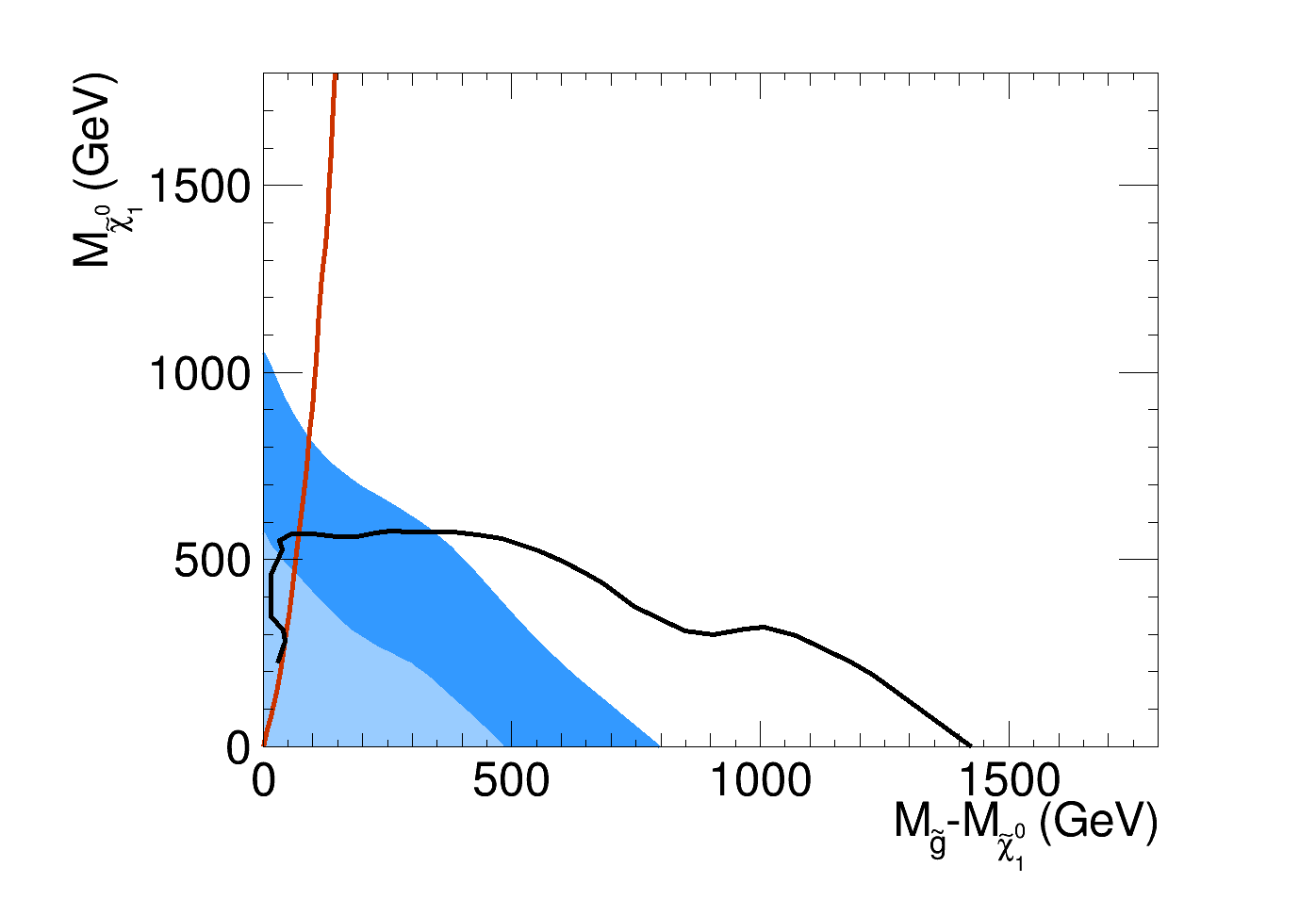}\\
\caption{Regions excluded by the monojet searches in the gluino-neutralino mass plane (upper panel) and in the mass splitting-neutralino mass plane (lower panel) by the 8 TeV run (light blue) and extrapolation for the 14 TeV run with 300 fb$^{-1}$ of data (dark blue). The black lines correspond to the ATLAS supersymmetric direct search limit, and the red lines to the relic density value as measured by Planck.\label{fig:mssm:gluino}}
\end{figure}%

The first MSSM simplified scenario we consider has $M_1$ and $M_3$ as main parameters, resulting in a pure bino neutralino and a gluino. The other masses are set to 40 TeV, apart from the stop sector parameters which are adjusted to obtain a light Higgs mass of 125 GeV. The gluino is assumed to decay exclusively to the lightest neutralino and two light quarks. The values of $M_1$ and $M_3$ are varied between 0 and 1.5~TeV. This scenario is of interest since it is well probed by the SUSY direct searches. Because the lightest neutralino is a pure bino, its interaction with matter is suppressed and it cannot be detected by direct dark matter detection experiments. Regarding the relic density, strong co-annihilations with the gluino are necessary to obtain a relic density in agreement with the Planck limits. Results are presented in Fig.~\ref{fig:mssm:gluino}, in the ($\tilde{g} - \tilde{\chi}^{0}_1$) mass plane for the 8 TeV run, as well as predictions for 14 TeV with 300 fb$^{-1}$ of data. The observed limit from the ATLAS Run~1 searches for direct gluino production in the jets + MET channel~\cite{Aad:2015iea} is also shown for comparison, as well as the region corresponding to the observed dark matter density. As can be seen, monojet searches are more constraining in the parameter region where the gluino and the lightest neutralino have almost degenerate masses, which also corresponds to the region where the relic density is close to or smaller than the observed value. The constraints from monojet searches on the neutralino mass can reach 600~GeV when this mass splitting is small. We observe that the monojet searches can marginally improve the constraints from the SUSY searches for mass splittings up to 50~GeV as the direct SUSY searches in this scenario are very strong. At the 14 TeV run, the monojet searches could probe neutralino masses up to 1.1~TeV.

\subsection{Degenerate squark scenario}

\begin{figure}[t!]
 \includegraphics[width=8.cm]{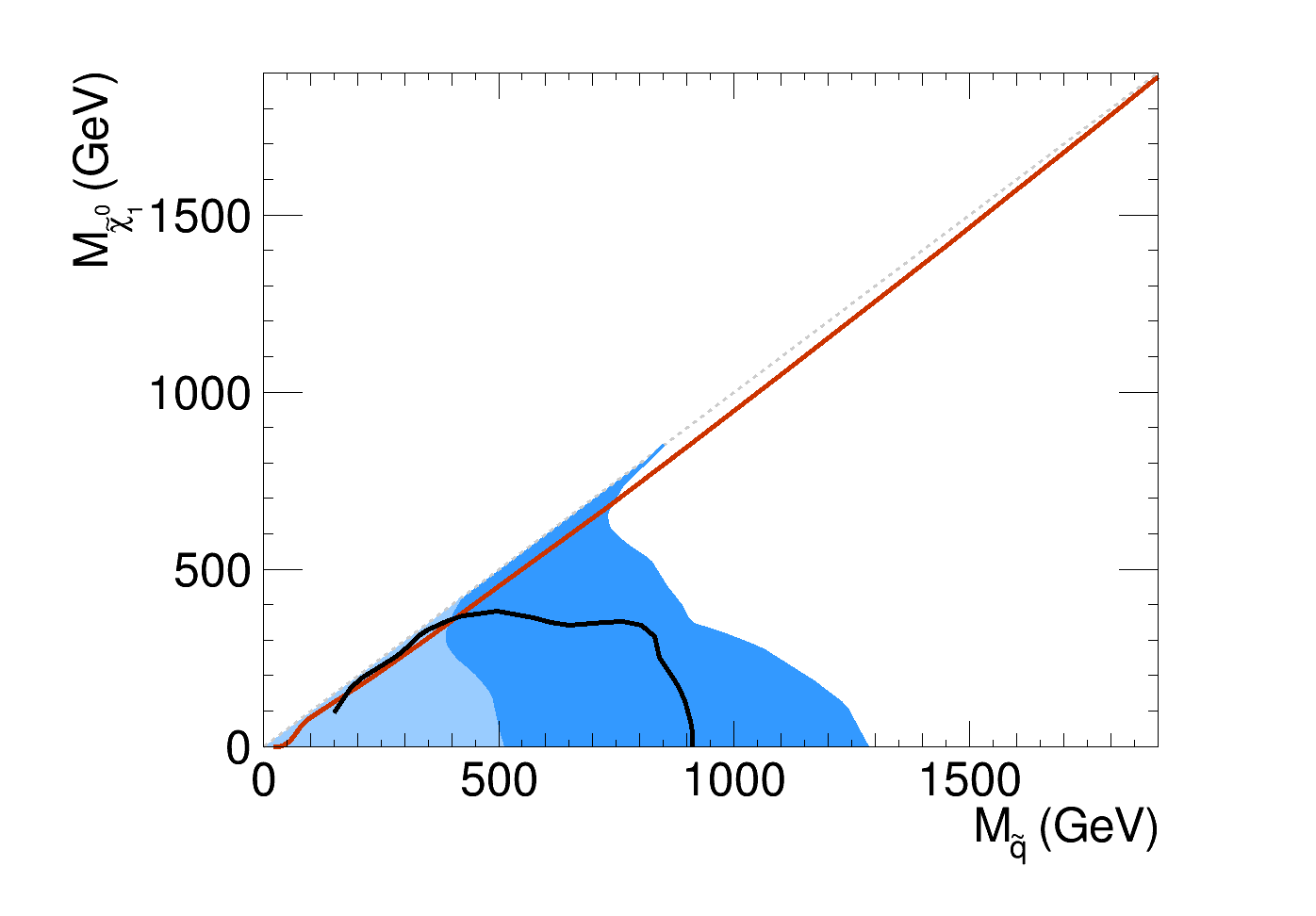}\\
 \includegraphics[width=8.cm]{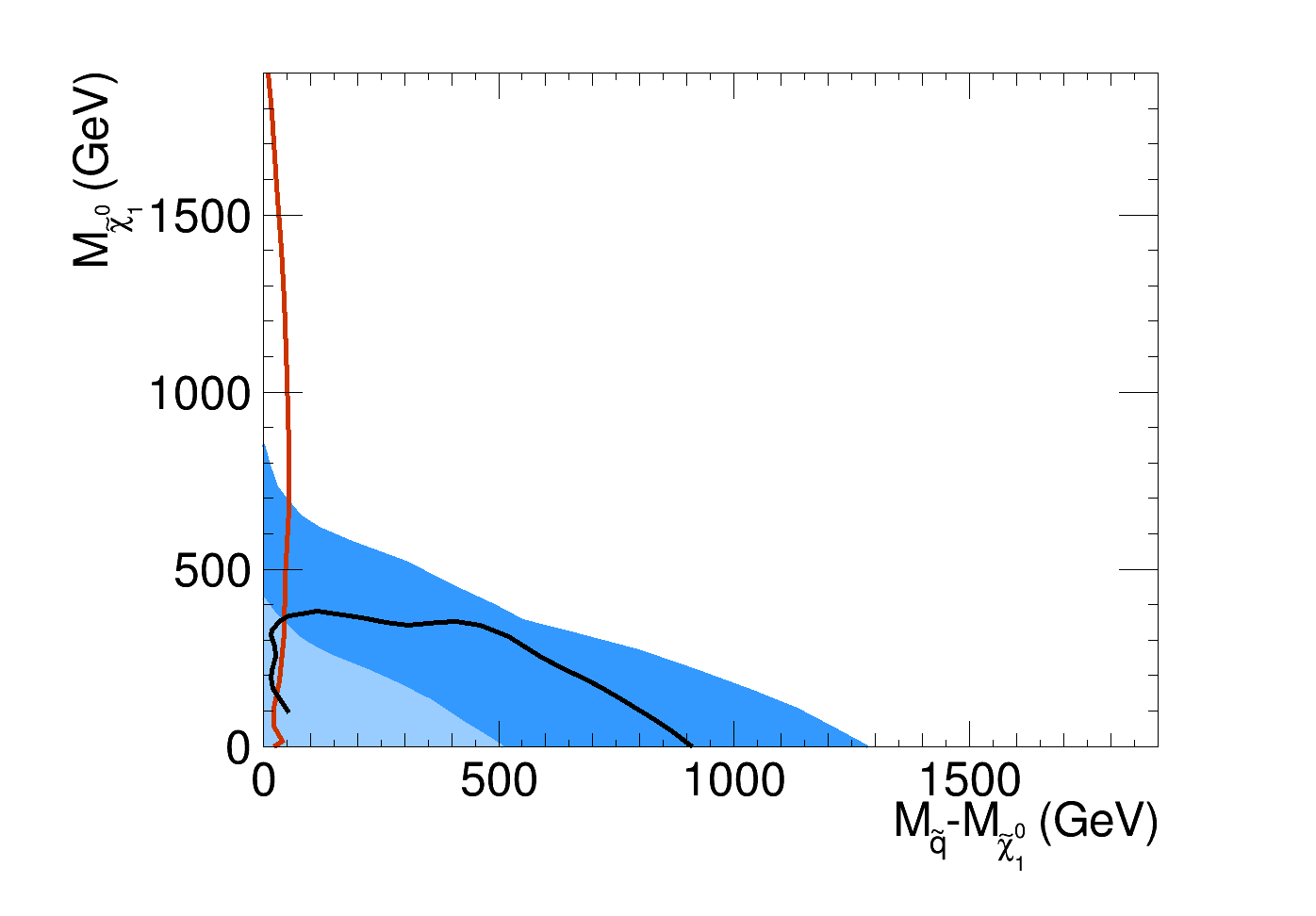}\\
\caption{Regions excluded by the monojet searches in the squark-neutralino mass plane (upper panel) and in the mass splitting-neutralino mass plane (lower panel) by the 8 TeV run (light blue) and extrapolation for the 14 TeV run with 300 fb$^{-1}$ of data (dark blue). The black lines correspond to the ATLAS supersymmetric direct search observed limit, and the red lines to the relic density value as measured by Planck.\label{fig:mssm:squark}}
\end{figure}

In the degenerate squark scenario the lightest neutralino is a pure bino, and all the eight first and second generation scalar quarks are taken to be light and degenerate in mass. The squarks decay exclusively to a quark and the lightest neutralino. This model has two parameters:
$M_1$ and the mass of the degenerate squarks, allowed to vary in the ranges $[0,2]$ TeV. This scenario is probed well by the LHC SUSY searches since all the eight first and second generation squarks participate to the cross sections. Again, since the neutralino is a bino, dark matter detection experiments are not sensitive enough to probe it, and a correct relic density requires a small mass splitting with the neutralino in order to have adequate co-annihilations. Results are shown in Fig.~\ref{fig:mssm:squark}, in the ($\tilde{q} - \tilde{\chi}^{0}_1$) mass plane. The limits from the ATLAS Run~1 direct squark searches in the jets + MET channel~\cite{Aad:2015iea} are also presented, in addition to the thin region corresponding to the Planck dark matter density. The most constrained region corresponds to that with the squarks and $\tilde{\chi}^{0}_1$ nearly degenerate in mass. In this region, the constraints on the $\tilde{\chi}^{0}_1$ mass go up to 400~GeV. Comparing the exclusions, we see that the monojet searches provide additional constraints to the direct SUSY searches in the region where the mass splitting is below 50 GeV. The 14 TeV run will probe neutralino masses close to 850 GeV.

\subsection{Light sbottom scenario}

\begin{figure}[t!]
 \includegraphics[width=8.cm]{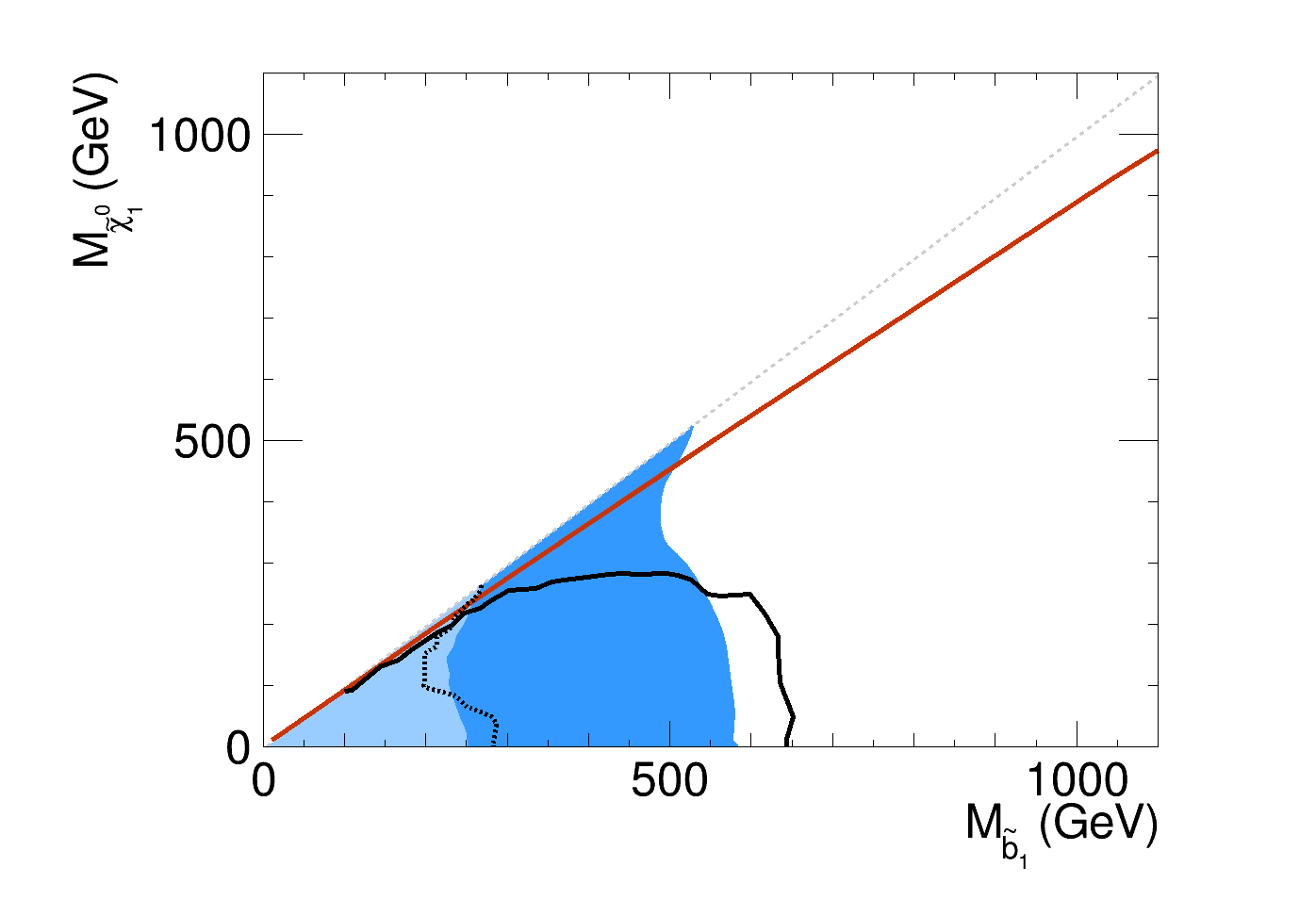}\\
 \includegraphics[width=8.cm]{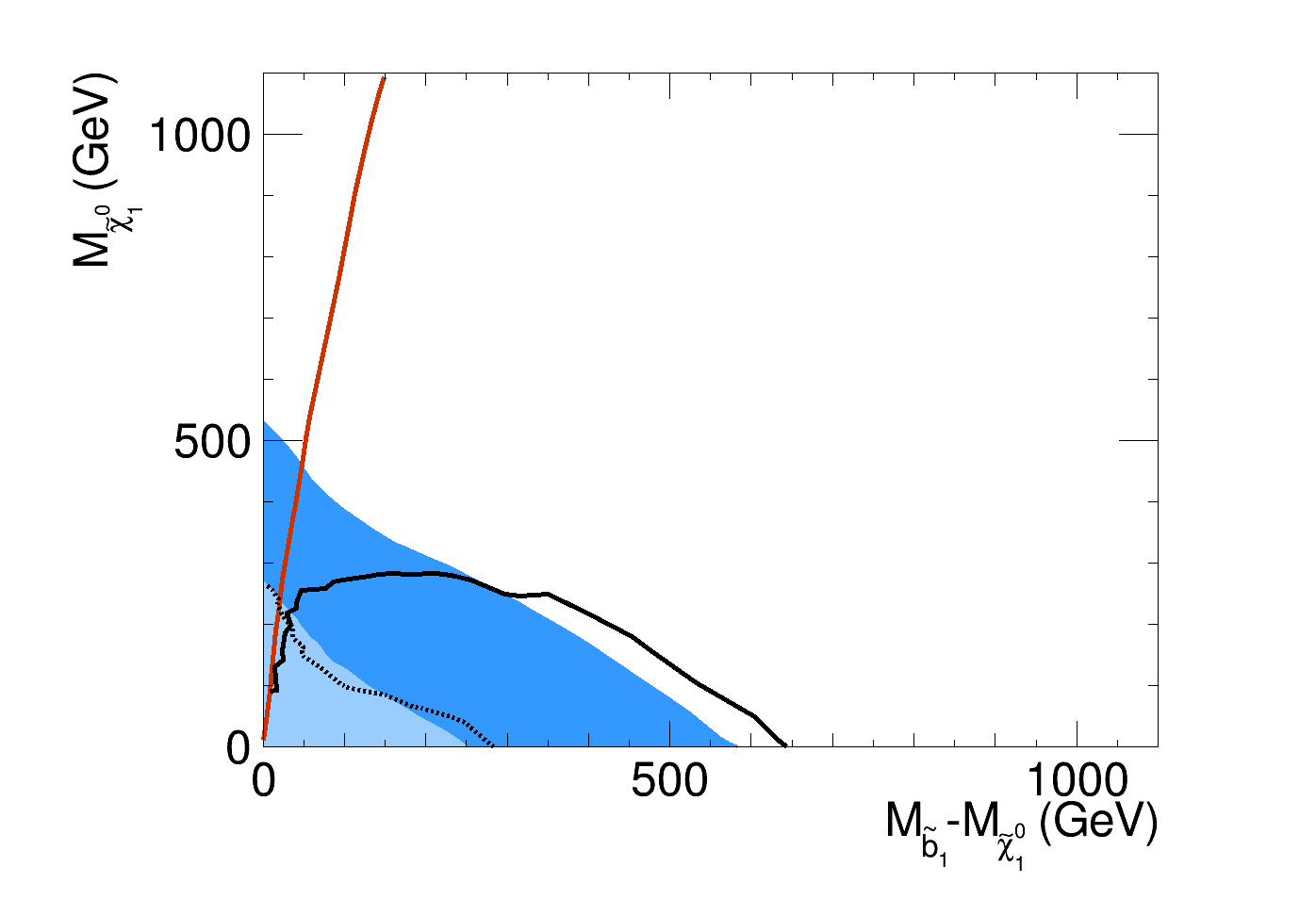}\\
\caption{Regions excluded by the monojet searches in the sbottom-neutralino mass plane (upper panel) and in the mass splitting-neutralino mass plane (lower panel) by the 8 TeV run (light blue) and extrapolation for the 14 TeV run with 300 fb$^{-1}$ of data (dark blue). The black solid lines correspond to the ATLAS supersymmetric direct search observed limit, the dotted lines to the ATLAS monojet search observed limit, and the red lines to the relic density value as measured by Planck.\label{fig:mssm:sbottom}}
\end{figure}%

The next scenario has a pure bino neutralino and a right-handed sbottom decaying exclusively to a bottom quark and a neutralino. The neutralino and sbottom masses are varied in the range $[0,1]$~TeV. The other masses are set to 40 TeV, apart from the stop sector where the parameters are adjusted to obtain a light Higgs mass of 125 GeV. Again, dark matter cannot be detected because of the elusive nature of the neutralino, and the correct dark matter density can be achieved thanks to co-annihilation with the sbottoms. Figure~\ref{fig:mssm:sbottom} summarises the results in the ($\tilde{b}_1 - \tilde{\chi}^{0}_1$) mass plane. The bounds from the ATLAS Run~1 sbottom searches in the 2~$b-$jets + MET and monojet channels~\cite{Aad:2015pfx} are also shown, as well as the region where the correct relic density is reached. We notice that the constraints on the $\tilde{\chi}^{0}_1$ mass goes beyond 250~GeV, while the direct sbottom searches probe neutralino masses up to 280~GeV, and the constraints are improved by the monojet searches for mass splittings below 40~GeV. As can be seen from the figure, our results at 8 TeV are rather similar to the observed ATLAS monojet search results, which can be considered as a validation of our analysis. At 14 TeV, neutralino masses up to 520 GeV can be probed.

\subsection{Light stop scenario}

We now consider a scenario with a wino-bino neutralino associated to a chargino close in mass, and a heavier scalar top quark. We define two separate regions: if the mass splitting of the squark with the chargino is smaller than the top mass, the scalar top decays exclusively to a bottom quark and the chargino, if the mass splitting is larger than the top mass, the stop decays exclusively to a top quark and the neutralino. The chargino subsequently decays to an off-shell $W$ boson and a neutralino, while the top quark can decay to an on-shell $W$ boson and a $b$ quark. The neutralino and stop masses vary in the range $[0,1]$~TeV. The other masses are set to 40 TeV, apart from the second stop. The parameters $M_1$ and $M_2$ are adjusted in order to obtain a correct relic density, following the relic density line described in the next subsection. The mass splitting with the stop is the only relevant parameter. This scenario is particularly interesting in the context of monojets because sizeable regions of its parameter space are not accessible to the standard stop searches. Regarding dark matter searches, we checked that scattering cross section of the neutralino with matter is one order of magnitude below the current experimental sensitivity. The neutralino-chargino set-up of this scenario is a specific case of the scenario described in the next subsection, so we refer the reader to the next subsection for more discussions about dark matter. Concerning the lightest Higgs mass, a correct value can be obtained by adjusting the stop 2 mass in the range 1--10 TeV and keeping no mixing in the stop sector.

\begin{figure}[t!]
 \includegraphics[width=8.cm]{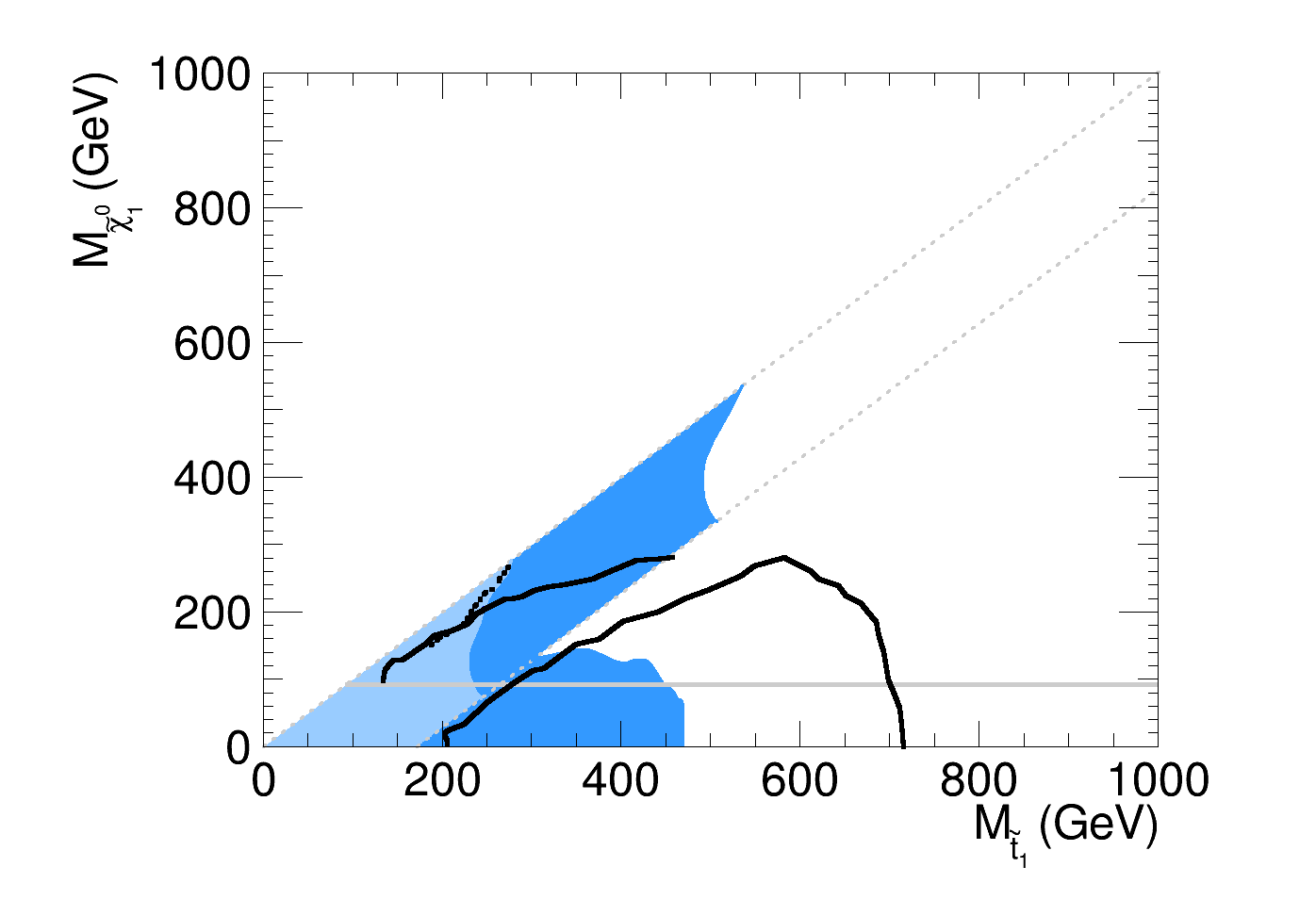}\\
 \includegraphics[width=8.cm]{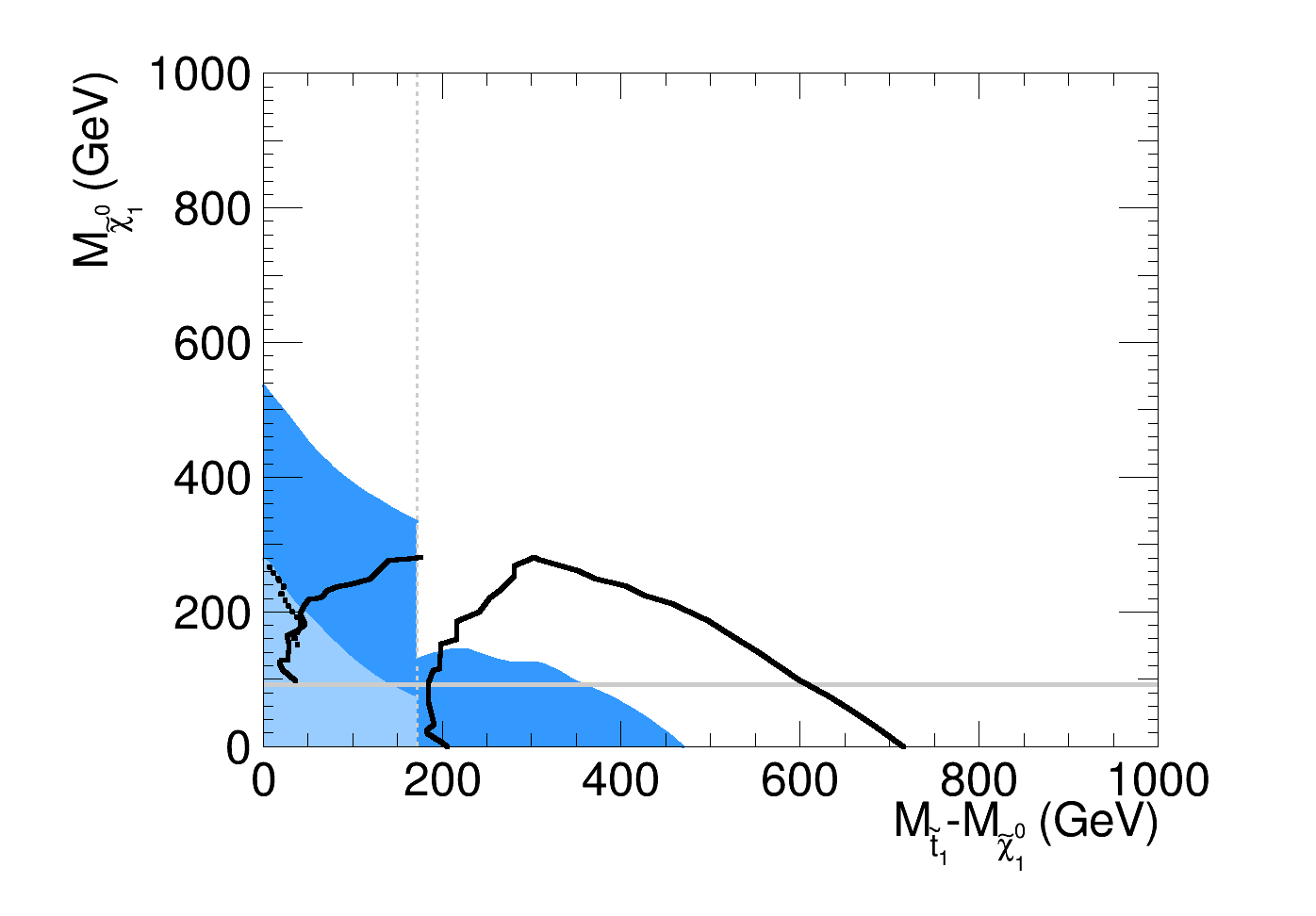}\\
\caption{Regions excluded by the monojet searches in the stop-neutralino mass plane (upper panel) and in the mass splitting-neutralino mass plane (lower panel) by the 8 TeV run (light blue) and extrapolation for the 14 TeV run with 300 fb$^{-1}$ of data (dark blue). The dotted lines correspond to a mass splitting equal to the top mass. On the left side of this line, the stop decays to a bottom and a chargino, and on the right side the stop decays to a top and a chargino. The black lines correspond to the ATLAS supersymmetric direct search observed limits. The horizontal gray lines correspond to the LEP chargino search limit.\label{fig:mssm:stop}}
\end{figure}

Results are shown in Fig.~\ref{fig:mssm:stop}, in the ($\tilde{t}_1 - \tilde{\chi}^{0}_1$) mass plane. The envelopes of the limits from the ATLAS Run~1 direct stop searches in the 2~top quarks + MET, 2~$b-$jets + MET, one isolated lepton + jets + MET, 2~leptons + MET, jets + MET and monojet channels~\cite{Aad:2015pfx}
are shown for comparison, for the two separate regions corresponding to mass splitting below and above the top mass. The monojet searches at 8 TeV only probe the region where the decay of the stop to a top and a neutralino is closed, corresponding to mass splitting below the top mass. In this region, the constraints are comparable to those published by the ATLAS collaboration. In the region where the stop can decay to a top and a neutralino, the monojet searches loose their sensitivity at 8 TeV since a top quark will manifest itself as an additional high-$p_T$ jet. The LEP constraints obtained in chargino searches \cite{LEP:chargino} are also shown for comparison. At 14 TeV, neutralino masses up to 550 GeV will be probed in the small mass splitting region, and the region where the stop can decay to the top quark and the neutralino will also be reached, so that mass splitting up to 450 GeV can be probed. For mass splittings above $m_t$, the monojet reach at 14 TeV is considerably worse than published 8 TeV stop search limits.

\subsection{Light neutralino and chargino scenario}
 
\begin{figure}[t!]
 \includegraphics[width=8.cm]{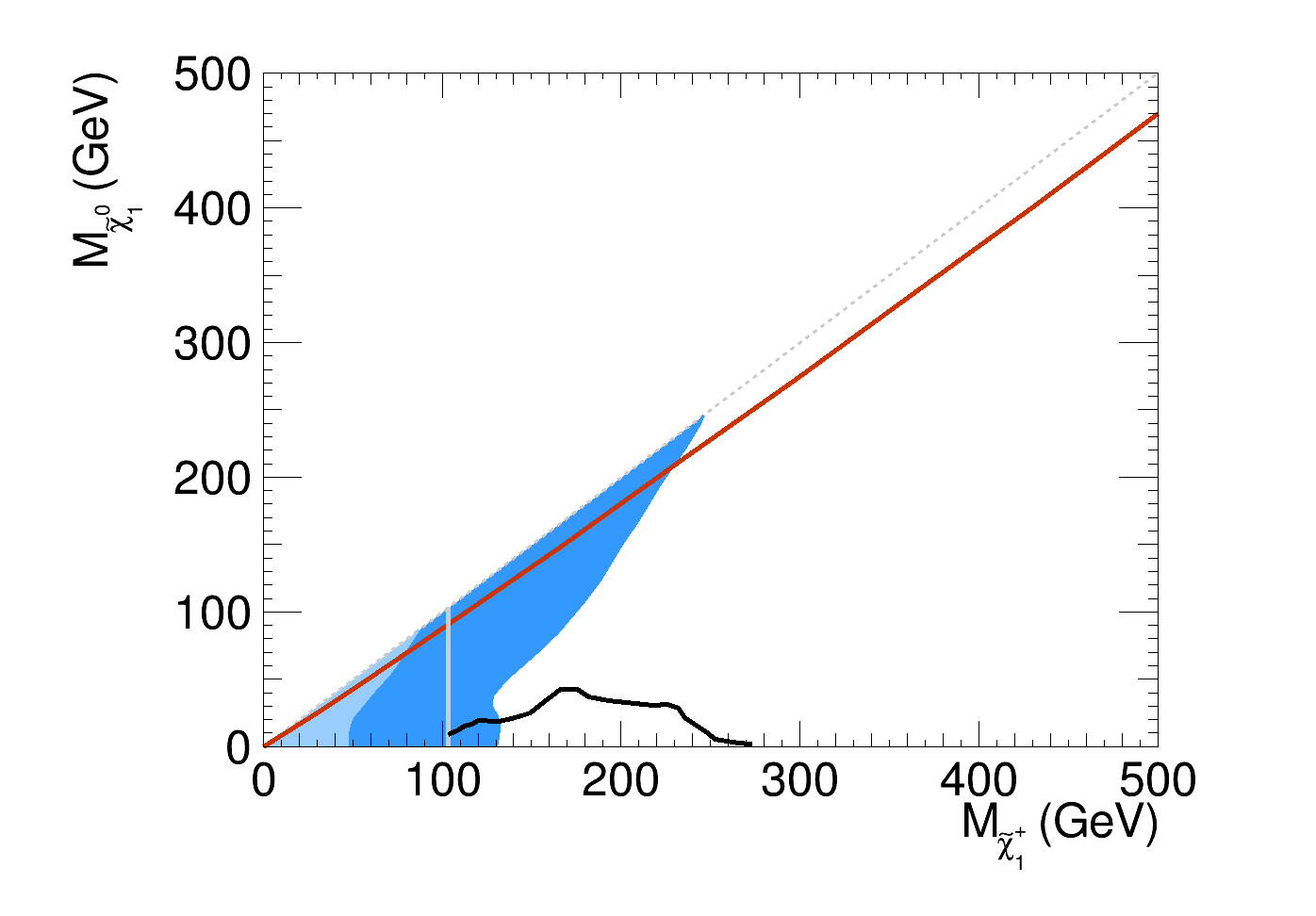}\\
 \includegraphics[width=8.cm]{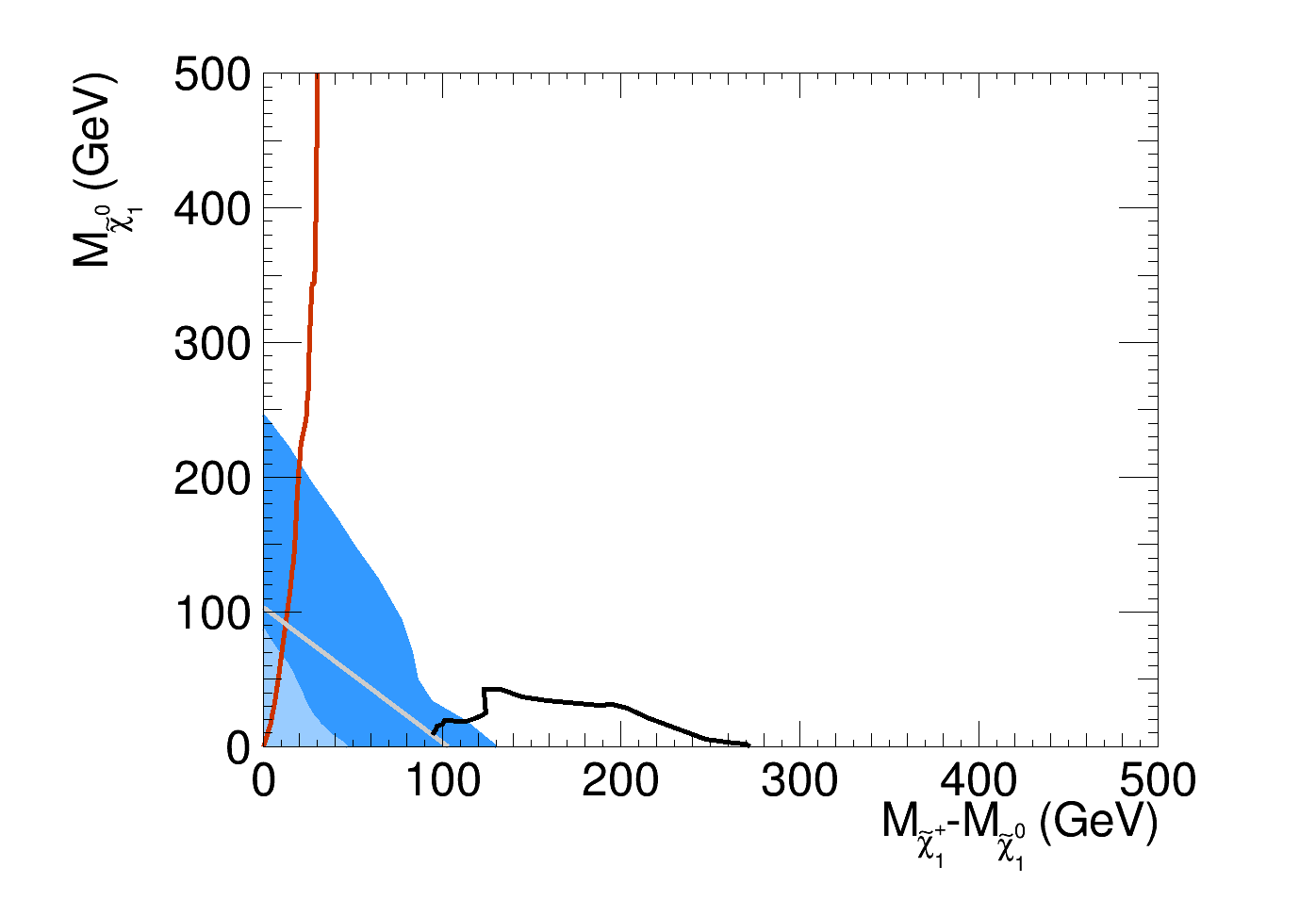}\\
\caption{Regions excluded by the monojet searches in the chargino-neutralino mass plane (upper panel) and in the mass splitting-neutralino mass plane (lower panel) by the 8 TeV run (light blue) and extrapolation for the 14 TeV run with 300 fb$^{-1}$ of data (dark blue). The black lines correspond to the ATLAS supersymmetric direct search observed limit, and the red lines to the relic density value as measured by Planck. The gray lines correspond to the LEP chargino search limit.\label{fig:mssm:neutralino}}
\end{figure}

The last simplified scenario we consider has $M_1$ and $M_2$ as the only low energy parameters. The other masses are set to 40 TeV, apart from the stop sector where the parameters are adjusted to obtain a light Higgs mass of 125 GeV. This scenario results in three light particles: two neutralinos, which can be bino-like, wino-like or bino-wino mixed states, and a wino chargino.
The $\tilde{\chi}^0_2$ is assumed to decay exclusively to the lightest neutralino and a (on- or off-shell) light Higgs boson $h$, and the chargino to the lightest neutralino and a (on- or off-shell) $W$ boson. The value of the $M_1$ and $M_2$ parameters are varied between 0 and 500~GeV. Results are shown in Fig.~\ref{fig:mssm:neutralino}, in the ($\tilde{\chi}^{\pm}_1 - \tilde{\chi}^0_1$) mass plane for the LHC 8 TeV run as well as the projection for the 14 TeV run with 300 fb$^{-1}$ of data. For comparison, the observed limit from the ATLAS Run~1 direct neutralino/chargino searches in the 2 and 3 leptons + MET and 1 lepton + $h$ + MET~\cite{Aad:2015eda} is also displayed.
The monojet search is particularly constraining in the region where the $\tilde{\chi}^{\pm}_1$ and $\tilde{\chi}^{0}_1$ have similar masses. In this region, the constraints on the neutralino mass can reach 90~GeV, and are complementary to the direct search limits. The monojet searches are particularly efficient in probing mass splittings below 40~GeV. The LEP constraints obtained in chargino searches \cite{LEP:chargino} are also shown for comparison, and they supersede the 8 TeV monojet search limits. At 14 TeV, the constraints will improve and neutralino masses up to 250 GeV can be reached, beyond the LEP limits.

\section{Monojet searches at higher energies}

\begin{figure*}[p]
 \includegraphics[width=8.cm]{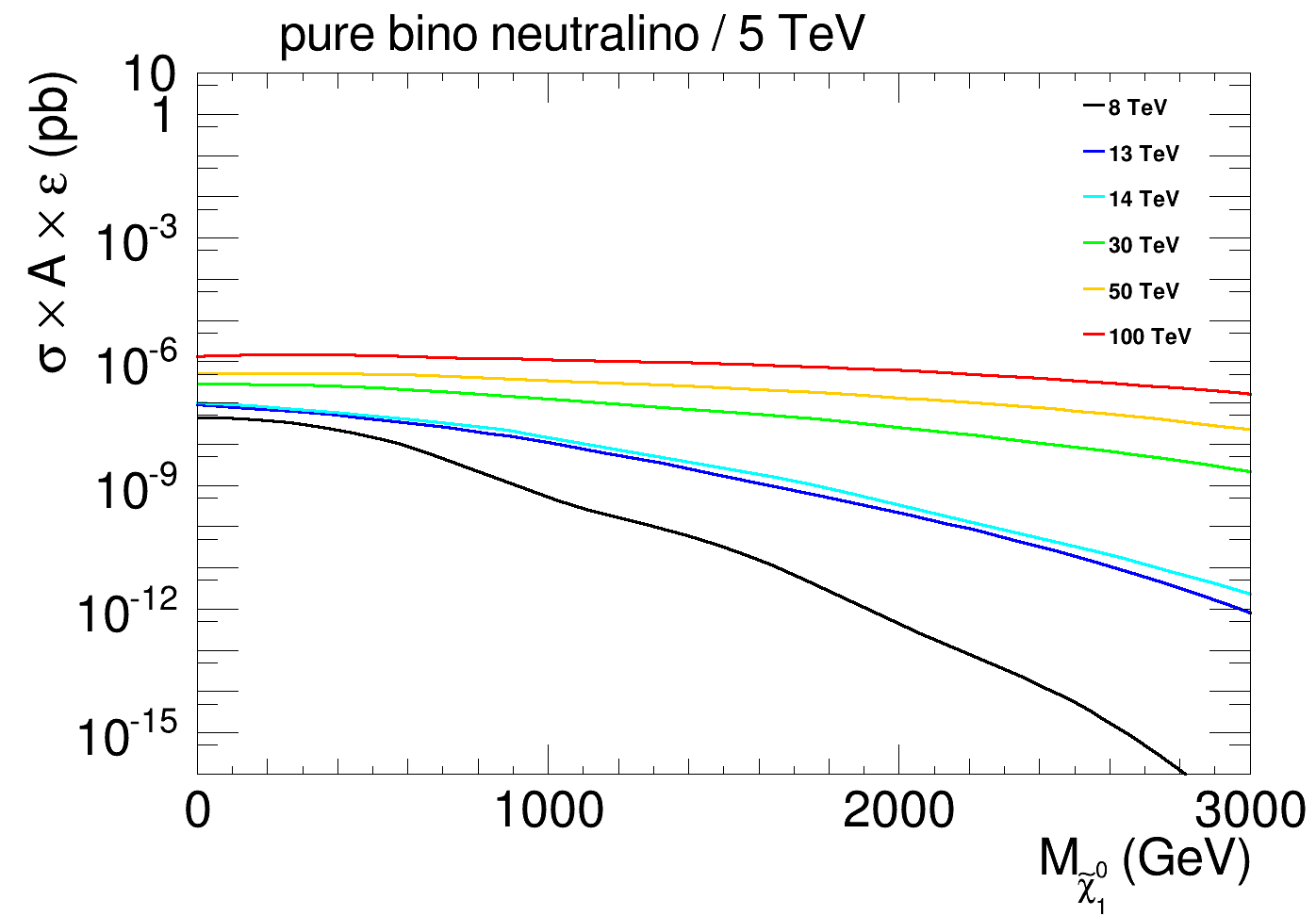}~
 \includegraphics[width=8.cm]{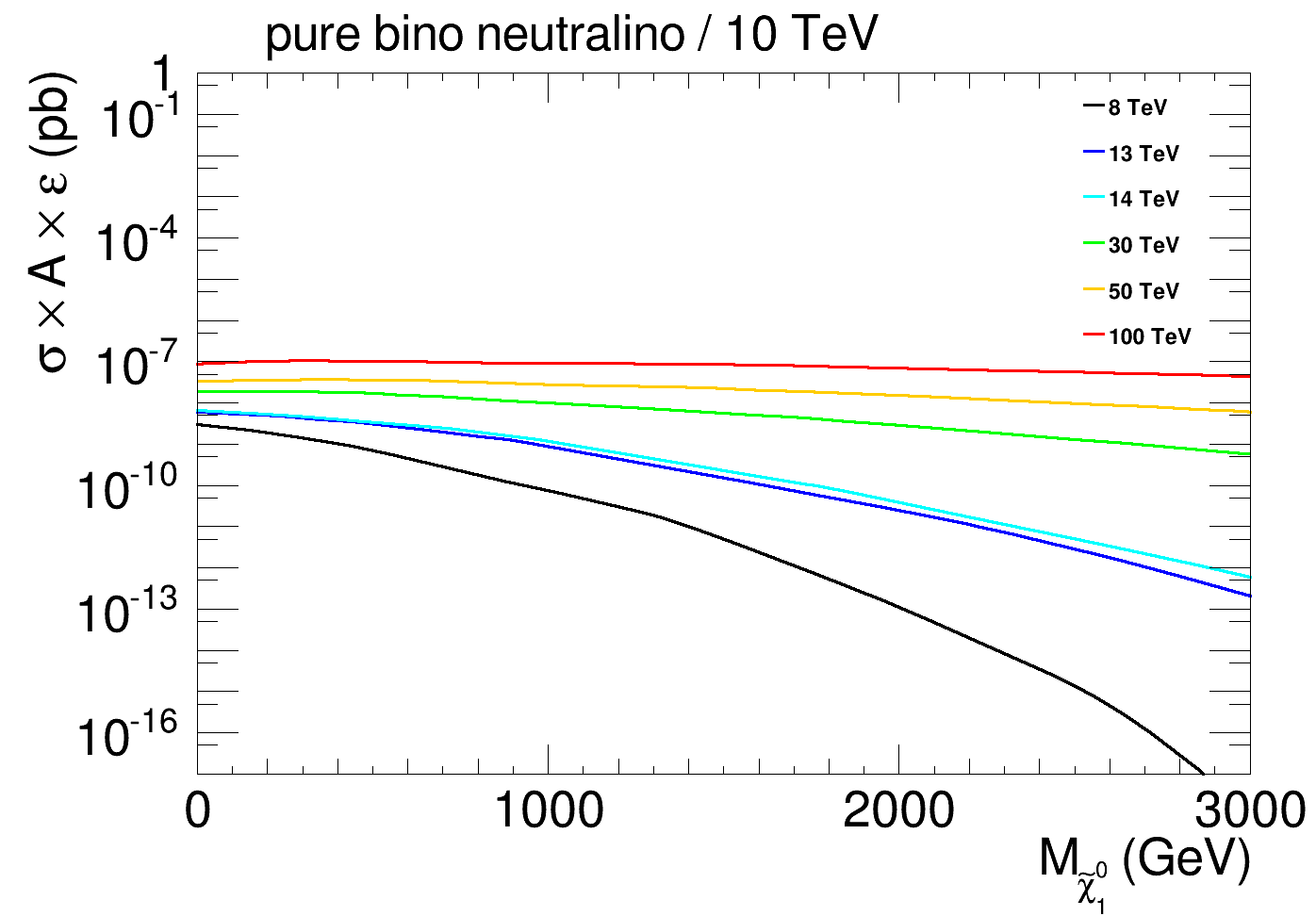}\\[-0.1cm]
 \includegraphics[width=8.cm]{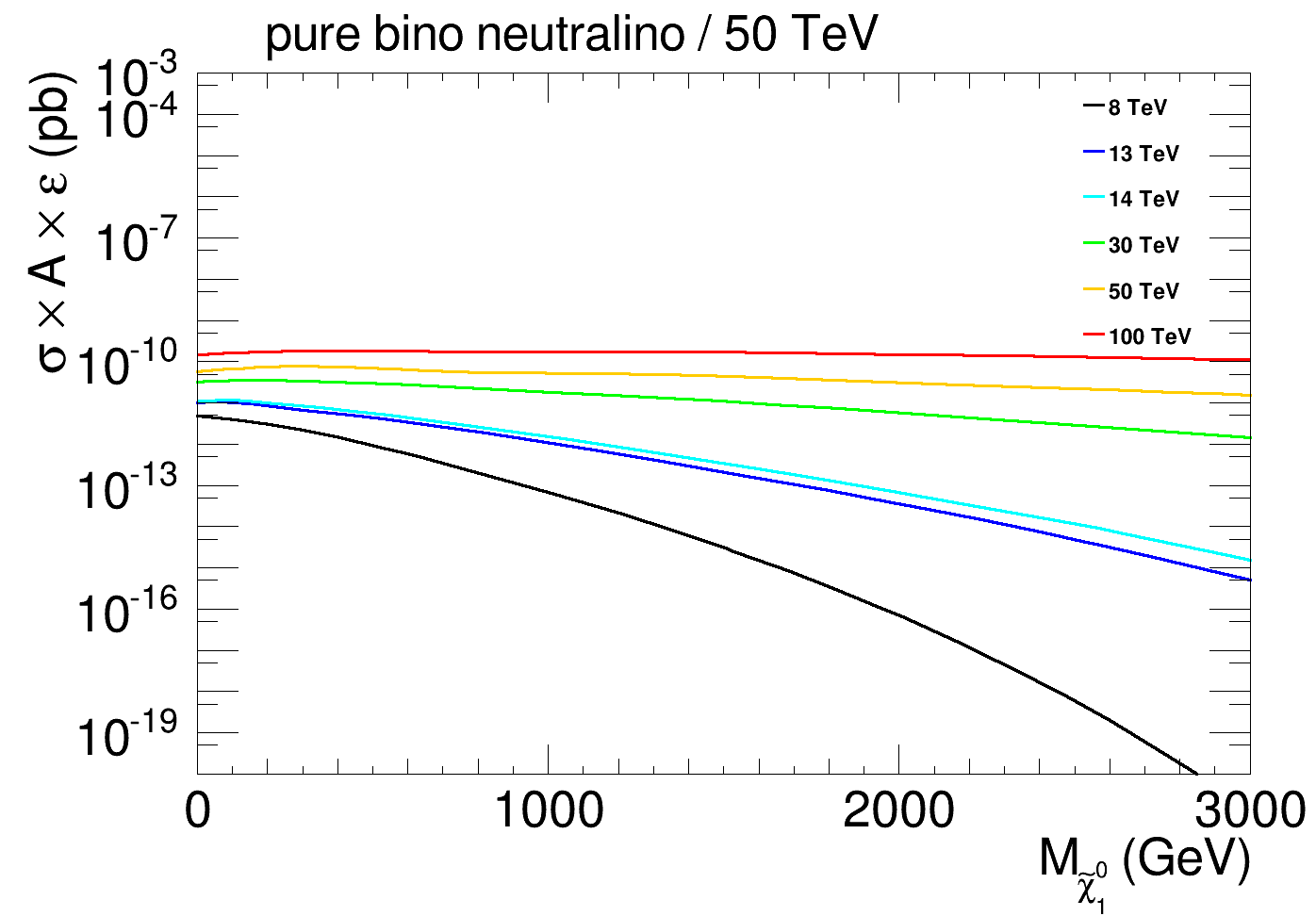}~
 \includegraphics[width=8.cm]{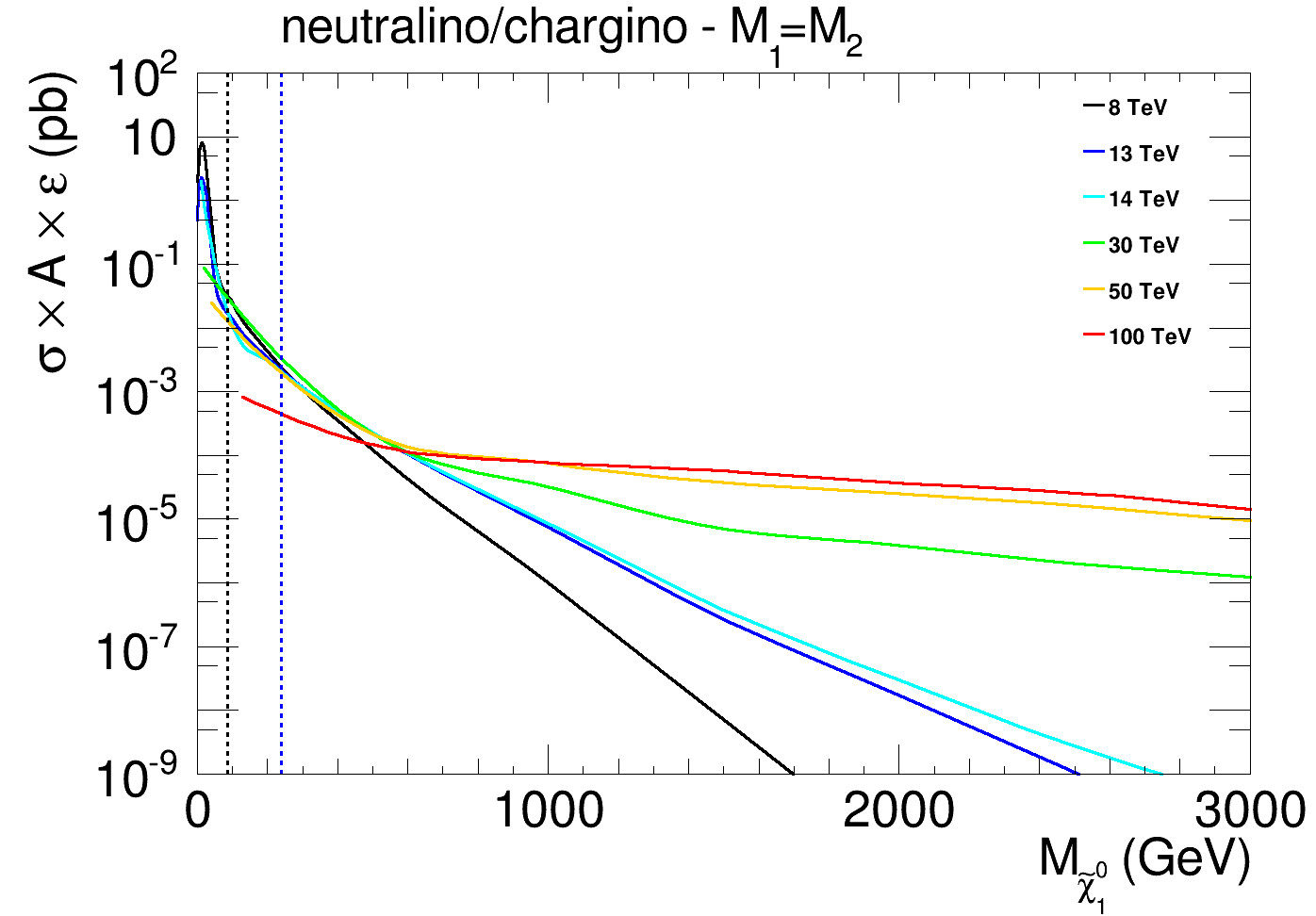}\\[-0.1cm]
 \includegraphics[width=8.cm]{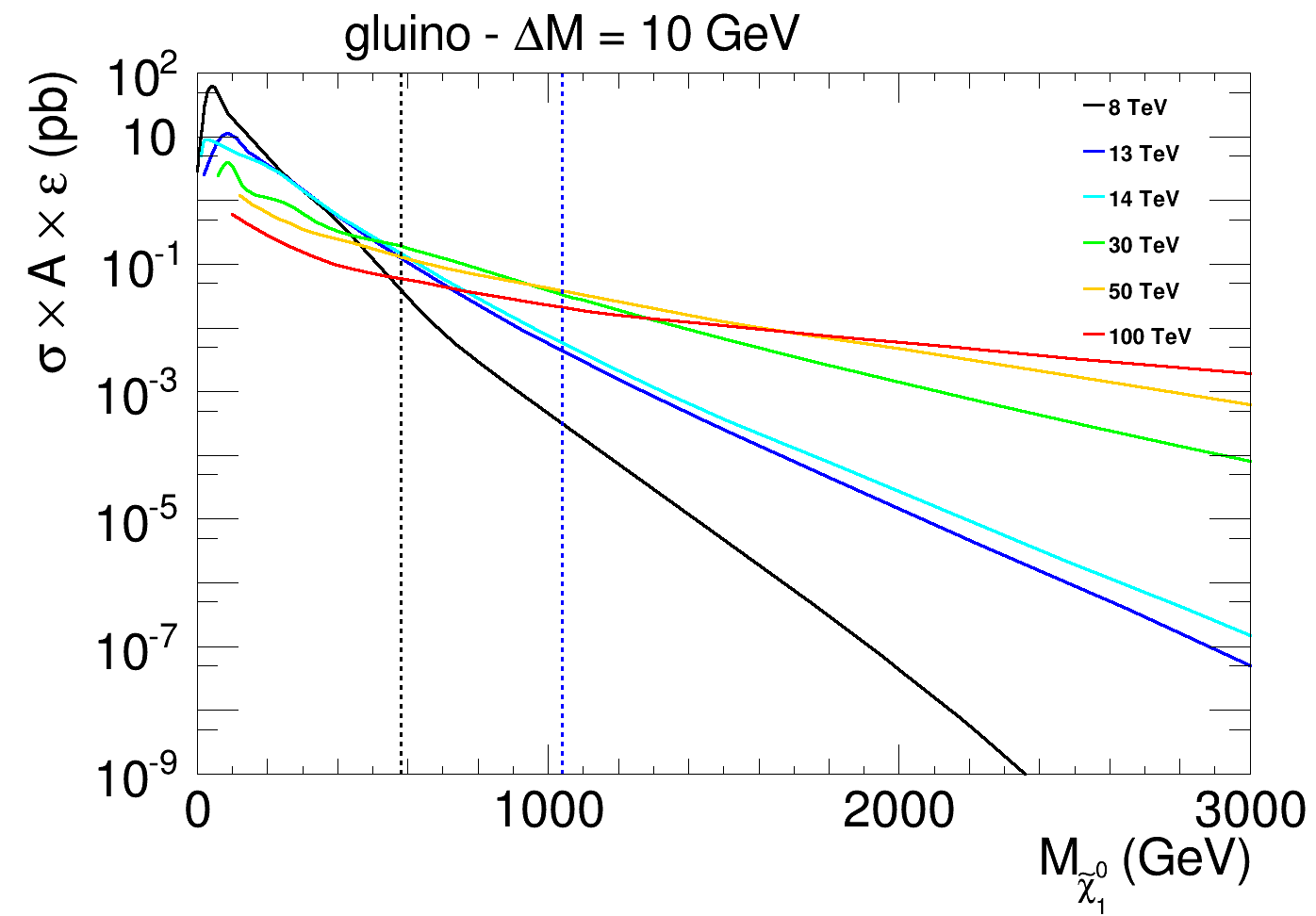}~
 \includegraphics[width=8.cm]{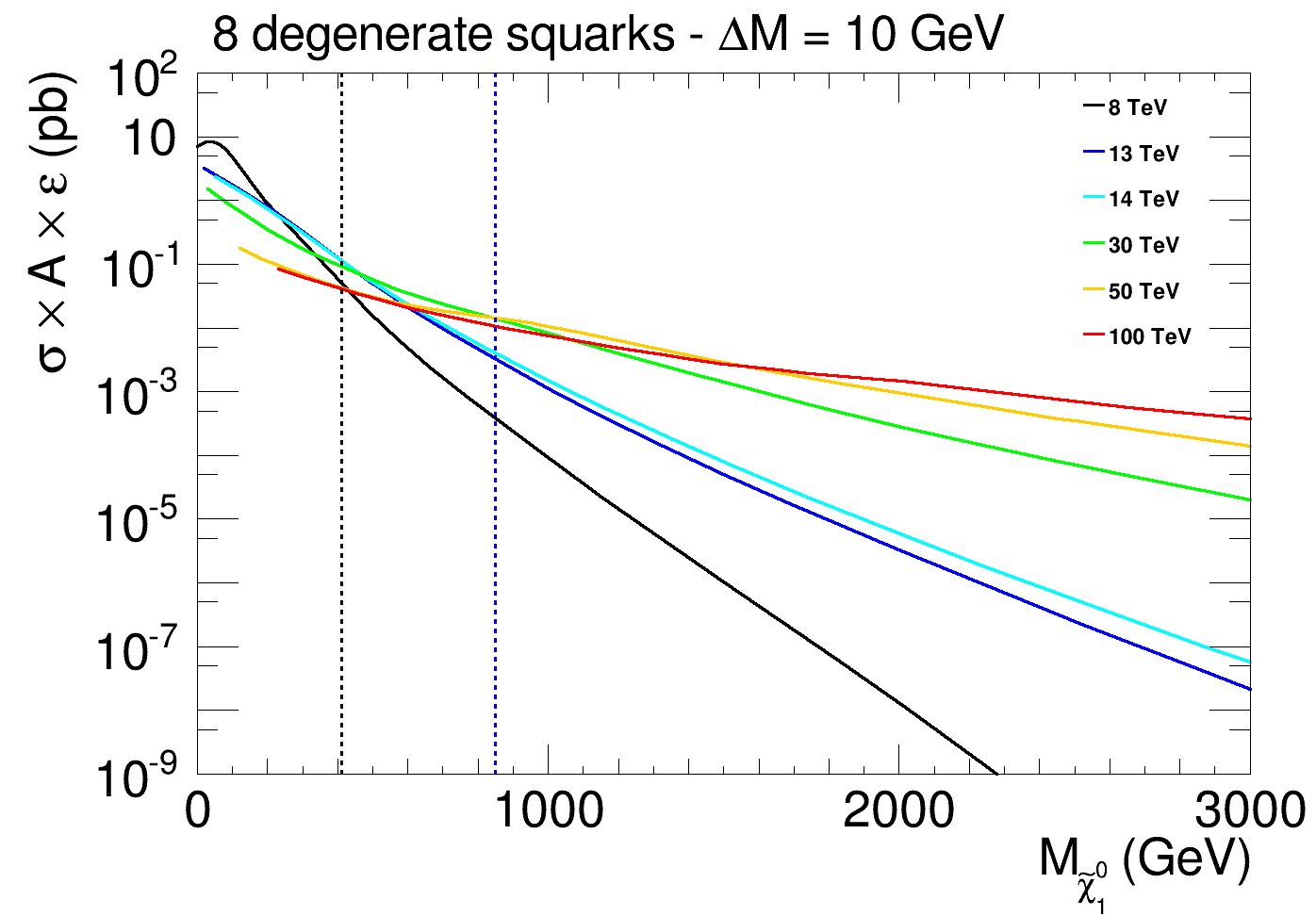}\\[-0.1cm]
 \includegraphics[width=8.cm]{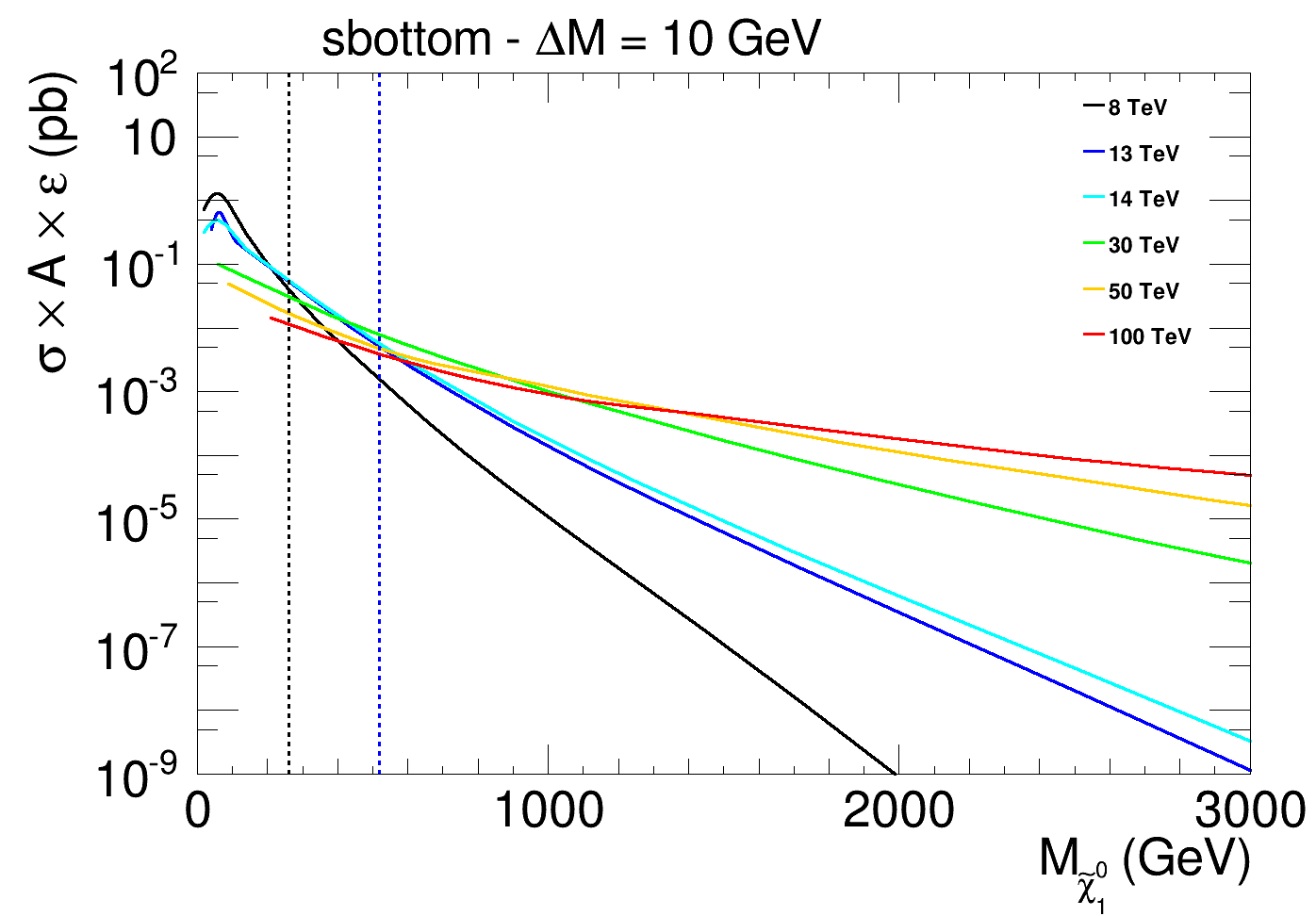}~
 \includegraphics[width=8.cm]{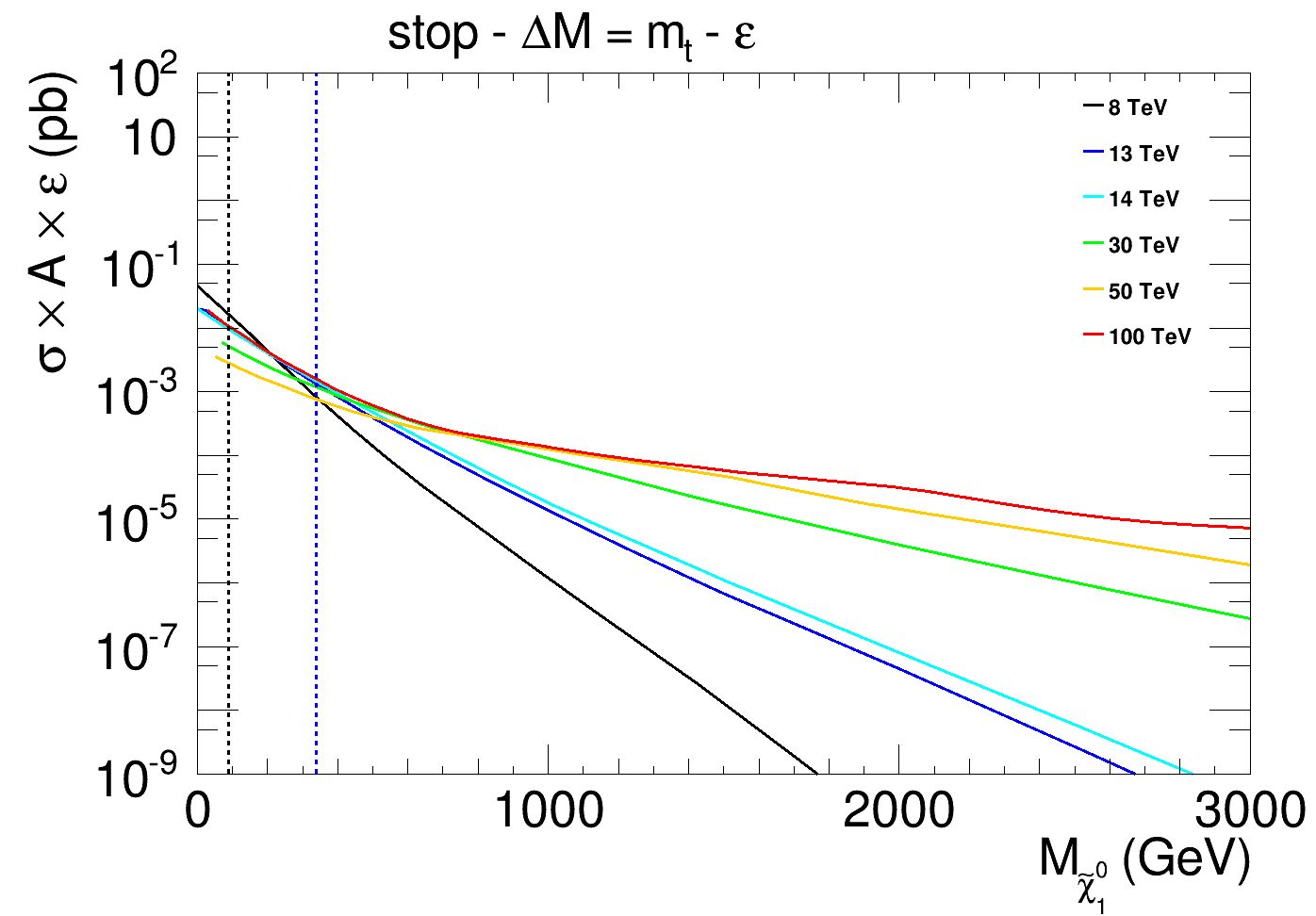}
\caption{Monojet production cross section times acceptance and efficiency as a function of the neutralino mass, for scenarios with a pure bino neutralino LSP with the other SUSY particles at 5 TeV  (upper left), at 10 TeV (upper right), at 50 TeV (upper middle left), a mixed wino/bino neutralino LSP and a chargino (upper middle right), a gluino (lower middle left), eight degenerate squarks (lower middle right), a sbottom (lower left) and a stop (lower right) with small mass splittings with the neutralino LSP. The different curves correspond to results at hadron colliders with $\sqrt{s} = 8$, 13, 14, 30, 50 and 100 TeV center of mass energies, imposing jet $p_T$ and missing $E_T$ cuts as discussed in the text. The black vertical dashed lines correspond to an indicative exclusion limit by the LHC Run~1, and the blue dashed lines to a prospective limit for the LHC 14 TeV run with 300 fb$^ {-1}$ of data.\label{fig:mssm:energy_dep}}
\end{figure*}
Monojet searches will remain a powerful tool for discovery at $pp$ colliders of increasing energy and luminosity. In order to assess the evolution of their sensitivity with energy and luminosity, we repeat our study for center of mass energies of 8, 13, 14, 30, 50 and 100~TeV for six different simplified MSSM models: a pure bino neutralino; a mixed state bino/wino in which there are two light neutralinos and one light chargino; the following cases with mass splitting of 10~GeV: a light gluino and a light bino neutralino, eight degenerate light squarks and a bino, a light sbottom and a bino; and finally a light stop and bino-wino neutralino and chargino with mass splitting slightly smaller than the top quark mass.
The mass splittings for the above-mentioned scenarios have been chosen to maximise the number of monojet events, but also to ensure the consistency between the SUSY model and the DM relic density constraints, requiring small mass splittings needed for co-annihilations (see for example \cite{deSimone:2014pda,Harigaya:2014dwa}). The calculation of the mass reach as a function of the luminosity and energy requires a detailed study accounting for the SM backgrounds, which goes beyond the scope of this paper. However, it is interesting to study the scaling of the product of the monojet production cross section times acceptance and efficiency as a function of the neutralino mass and the collider energy. The acceptance is defined by $\sqrt{s}$-dependent lower cuts on the jet $p_T$ and missing $E_T$, scaled from typical values adopted in the 8 TeV searches and given by $\sqrt{s}/(8$ TeV)$\times 250$ GeV. The results are shown in Fig.~\ref{fig:mssm:energy_dep}. For the bino case, we do vary the mass of the other SUSY particles between 5 and 50 TeV since the monojet cross section is sensitive to it.

The limits obtained for 8~TeV give the current status of these searches and their extrapolation to 300 fb$^{-1}$ of data for the LHC 14 TeV run are given for comparison. Although the change in cross section times efficiency from 8 to 14 TeV as a function of the mass is relatively small, the increase in mass coverage afforded by 14 TeV is very significant.

This motivates a possible increase of the energy up to 30~TeV, in principle compatible with the radius of the LHC tunnel and dipoles of new technology, and beyond. The pure bino case remains out of reach due to its small cross section but a collider with an energy at the order of 100~TeV and high luminosity would possibly provide enough statistics for probing neutralino masses in all the other scenarios up to more than 3~TeV. This upper limit is particularly interesting since a relic density compatible with the CMB data can be reached for wino and higgsino neutralinos of masses between 1 and 3~TeV in absence of co-annihilations with sfermions, a window which could tantalisingly be accessible at a 100~TeV collider.

\section{Conclusion}

The search for monojets is a powerful tool to explore new processes at hadron colliders. 
To illustrate this in a quantitative way, we have considered simplified MSSM scenarios in which only a few relevant degrees of freedom are considered.
We showed that direct searches in the jets/leptons + MET final states and monojets are highly complementary, the latter improving the sensitivity in regions with small mass splittings. Such regions are highlighted by DM relic density involving co-annihilation processes. Recasting the monojet searches in the MSSM, it is important to consider all the relevant topologies, namely processes involving squarks and gluinos escaping the detection in addition to the usual WIMP-WIMP-jet topologies, as the former result in large cross sections at the LHC and can be dominant when the squark/gluino mass becomes nearly degenerate with the lightest neutralino.
We find that the complementarity of the monojet and direct SUSY searches is particularly striking in the case of the light neutralino and chargino scenario. Small mass splittings in the gaugino sector naturally arise when the lightest neutralino is a pure wino or higgsino, resulting in weaker constraints from direct multi-lepton + MET searches but increased sensitivity for the monojets.

\subsection*{Acknowledgements}

A.A. and F.M. acknowledge partial support from the European Union FP7 ITN INVISIBLES (Marie Curie Actions, PITN-GA-2011-289442). The work of M.B. was supported in part by the U.S. Department of Energy grant number DE-SC0010107.


\end{document}